\def \yskip{\penalty-50\vskip3pt plus 3pt minus 2pt}
\def \reference{\par \yskip \noindent \hangindent .4in \hangafter 1}
\def \abc#1#2#3#4 {\reference#1, {\sl#2}, {\bf#3}, #4}
\def \blank {\lower 5pt\hbox to 0.75in{\hrulefill}}

\def \s{~\rm{s}}
\def \km{~\rm{km}}

\def \AU{~\rm{AU}}
\def \yrs{~\rm{yrs}}
\def \yr{~\rm{yr}}

\def \kpc{~\rm{kpc}}
\def \lae{\mathrel{<\kern-1.0em\lower0.9ex\hbox{$\sim$}}}
\def \gae{\mathrel{>\kern-1.0em\lower0.9ex\hbox{$\sim$}}}
\documentstyle[11pt,aasms4]{article}

\begin{document}
\small

\setcounter{page}{1}
\begin{center} \bf 
VISUAL WIDE BINARIES AND THE STRUCTURE \\
OF PLANETARY NEBULAE
\end{center}

\begin{center}
Noam Soker \\
Department of Physics, University of Haifa at Oranim\\
Oranim, Tivon 36006, ISRAEL \\
soker@physics.technion.ac.il 
\end{center}

\begin{center}
\bf ABSTRACT
\end{center}

 In a recent work Ciardullo {\it et al.} (1999) list 19 planetary nebulae
surveyed by the Hubble Space Telescope for the presence of resolved
visual binary companions of their central stars. 
 For ten planetary nebulae they argue for probable physical association
of the resolved stars with the central stars, while for nine the association
is less likely.
 Such stellar companions, at orbital separations of hundreds to thousands
of astronomical units, will cause the structures of these planetary nebulae
to possess a non-axisymmetrical signatures.
 By using images from the literature of these 19 planetary nebulae,
I demonstrate that the structures of the planetary nebulae are compatible
in most cases with Ciardullo {\it et al.}'s arguments for an association,
or not, of the resolved stars with the planetary nebulae central stars.
 This shows that the departure, or not, of a planetary nebula from
having pure axisymmetrical structure can be used to strengthen an
argument for an association, or a non-association, of a putative wide
companion with the stellar progenitor of the nebula.

\noindent 
{\it Subject heading:}   
Planetary nebulae:general
$-$ binaries: visual
$-$ stars: AGB and post-AGB


\section{INTRODUCTION}

 In a recent paper Ciardullo {\it et al.} (1999) resolved the
companions (or possible companions) of 19 planetary nebulae (PNs) nuclei.
 By assuming that the resolved companions are main sequence stars, they 
calculated the distances to these PNs.
 The projected orbital separations of the binary nuclei are in the range
of a few hundred to several thousand astronomical units.
 At such large orbital separations the companions will not influence the
mass loss from the PN progenitors.
 However, the companions can influence the structures of the descendant PNs.
For orbital periods of $P_{\rm orb} \lesssim 10 \tau_{PN}$,
where $\tau_{PN}$ is the formation time for the nebula, the orbital motion
may cause a departure from axisymmetrical structure
(Soker 1994), while for larger separations the companion may blow a small
bubble inside the nebula (Soker 1996).
 Such a bubble, formed by a very wide companion, might be the
``vertical bridge'' observed by Corradi {\it et al.} (1999) near a star
located in Wray 17-1.

 The list of binary companions presented by Ciardullo {\it et al.} allows
a qualitative comparison between the properties of the binary nuclei
and the observed structures of these PNs.
 This is the aim of this paper: to search for a consistency between the
claimed binary nuclei and the PNs morphologies, thereby to support
the arguments of Ciardullo {\it et al.} (1999) for an association, or not,
between the observed companions and the central stars of the PNs. 
 As an example, let us examine the Stingray nebula, for which a visual
companion was resolved by the Hubble space telescope near its central
star (Bobrowsky {\it et al.} 1998).
 The projected orbital separation is $\sim 2200 \AU$, while the
outer nebular radius is $\sim 8000 \AU$.
 This radius implies a nebular expansion age of $\tau _{PN} \sim 4000$ years
for an expansion velocity of $10 \km \s^{-1}$.
 A companion at such a large separation cannot be
responsible for the axisymmetrical structure of this PN.
 Its orbital separation, which implies an orbital period of $\sim 10^5 \yrs$,
may manifest itself in a departure from axisymmetry.
 During the time of $\tau_{PN} \simeq 4000$, the mass-losing
star would have moved a distance of
$\sim 0.5(4000/10^5)\times 2 \pi \times 2200 \AU \simeq 275 \AU$ in its
orbital motion (assuming the two stars have equal masses).
 This is $ \sim 3 \%$ of the nebular radius, and it implies a clear
deviation from axisymmetry.
 This PN, indeed, has a large degree of deviation from axisymmetry,
part of which may be attributed to the presence of this companion.
 However, this PN has a complicated structure, including a dense ring
which is inclined to the long axis of the nebula.
 The overall degree of deviation from axisymmetry seems to me to be too
large to be solely explained by the visual companion, and I hypothesize
that a closer companion in an eccentric orbit may exist. 
 The visually resolved companion's wind may also have shaped the nebula
in its immediate vicinity.
 The Stingray nebula deserves further observations. 

 The paper is organized as follows.
 In $\S 2$ I find the masses of the main sequence companions, and
calculate the orbital periods and velocities.
 I then mark each PN according to whether its structure is compatible
or not with the properties of its central binary system, and
in $\S 3$ I elaborate on the individual PNs.
{{{  In $\S 4$ I compare the sample studied in $\S 3$ to the sample of
PNs for which Ciardullo {\it et al.} found no companions. }}}
 A short summary is in $\S 5$.

\section{PROPERTIES OF BINARY NUCLEI}

 In Table 1 the same 19 PNs that are given by Ciardullo {\it et al.}
(1999) in their table 7 are listed.
 The columns in Table 1 are:
(1) The common PN name.
(2) The PN G designation (galactic coordinates).
(3) The secondary mass in $M_\odot$.
     It is calculated from the magnitude $M_V$, which is found from the
     colors (V-I) and reddening [E(B-V)] as given by Ciardullo {\it et al.}
     Only for K 1-27 was the companion's mass taken to be $0.6 M_\odot$
     since Ciardullo {\it et al.} argue that the companion is a WD
     rather than a main sequence star. 
(4) The projected orbital separation in AU. The values are taken
    from Ciardullo {\it et al.}, where more details can be found. 
(5) The orbital period, in years, calculated by taking the progenitor
     of the PN (now the WD) to be of mass $0.8 M_\odot$.
     The real orbital separation is likely to be larger
     than the projected orbital separation.
     An effect in the opposite direction is the mass-loss process during
     the AGB phase.
     The orbital separation increases during this phase, and hence it was
     smaller at the beginning of the process.
     The projection effect is likely to dominate in most cases, hence the
     orbital periods are likely to be longer than those given in the table.  
(6) The orbital velocity of the primary around the binary center of mass,
    taking the projected orbital separation and assuming a circular
    orbit and a mass of $0.8 M_\odot$ for the WD progenitor.
(7) The nebular diameter as given by Ciardullo {\it et al.} (1999), in
    parsecs.
(8) The classification of Ciardullo {\it et al.} (1999; Table 7):
    P$-$Probable binary; S$-$Possible binary; D$-$Doubtful.
(9) Departure from axisymmetry: Y$-$a departure from axisymmetry is
    clearly seen; N$-$no clear departure from axisymmetry is seen in the
    images.
    By departure from axisymmetry, I refer only to large-scale departures,
    and not to small-scale blobs, filaments etc.
(10) My estimate of the compatibility of the PN departure from axisymmetry
    and the possibility that its binary companion caused this departure:
    ($+$)Compatible; $(?)$hard to tell.
  For example, for A 30 no departure is expected since the orbital period
is very long, and indeed no large-scale departure from axisymmetry 
is observed, hence the $+$ sign.
 For NGC 7008 a deviation from symmetry is expected and is indeed observed,
hence a $+$ sign.
 In K 1-14 a departure is observed, but the main signature is on the
outskirts of the nebula, and hence an interaction with the ISM is
possible.
  If the companion is associated with the PN central star, as claimed by
Ciardullo {\it et al.}, then it can also cause a departure.
 Hence for this PN I put a question mark, since it is not clear if the
morphology is compatible with the claimed companion, or it is
solely due to an interaction with the ISM.

 Four main processes can cause a large-scale deviation from axisymmetry.
($i$) Interaction with the ISM.
In this case the most prominent features are on the outskirts on the
nebula (e.g., Tweedy \& Kwitter 1994, 1996), with smaller, or none at all,
deviations from axisymmetry in the inner regions of the nebula.
 It is worth noting that in large PNs the ISM may penetrate the
outer regions of the nebula, and influence the inner structure as
well as the outer regions (Dgani \& Soker 1998). 
($ii$) A close companion in an eccentric orbit (Soker, Rappaport, \& Harpaz
1998).
 This occurs when the companion is close enough to influence the mass
loss process from the AGB star, and the eccentricity is large.
($iii$) A wide companion, with an orbital period in the range of
$0.3 \tau_{PN} \lesssim P_{\rm orb} \lesssim 30 \tau_{PN}$, where
$\tau_{PN}$ is the formation time of the relevant part of the nebula.
The formation time can be $\tau_{PN} \sim {\rm several} \times 10^4 \yr$
for a halo,  $\tau_{PN} \sim {\rm several} \times 10^3 \yrs$ for the
dense inner shell, and $\tau_{PN} \simeq {\rm several} \times 100 \yrs$
for possible jets.
 For most of the PNs listed in Table 1, the relevant time is the expansion
time of the nebula to its current radius
$\tau_{PN} \sim R_{PN}/10 \km \s ^{-1} \sim 10^4- 5 \times 10^4 \yrs$.
  Another requirement of the latter mechanism is that the velocity of the
mass-losing star around the center of mass $v_1$ will be not too small.
 This means a companion of mass $M_2 \gtrsim 0.3 M_\odot$, depending on
the orbital separation (see $\S 4$).
($iv$)
 During the mass loss process itself, if one or a few
long live cool (or hot) spots exist on the surface of the AGB star.
 This process seems to be important to massive stars
(e.g., as suggested for the $\sim 30 M_\odot$ star HD 179821
by Jura \& Werner 1999), but it is not clear if it can operate
efficiently in AGB stars, where the strong convection may
not allow such spots to live long enough.

 Simple estimates suggest that the deviation from axisymmetry,
e.g., the dislocation of the central star from the center of the nebula,
will be of the order of $v_1/v_e$, where $v_e$ is the expansion velocity
of the nebula.
 However, interaction of the wind with previously ejected mass from the
AGB star may change the density of the wind in a different sense on both
sides, and because of the sensitivity of emission to the density,
a larger degree of departure from axisymmetry will be observed.
 The observed degree can be lower than expected if the direction of
deviation is not perpendicular to the line of sight;
 if it is along the line of sight, the departure from axisymmetry will not
be observed.
 In most cases I expect the degree of axisymmetry  to be in
the range from one to several times $v_1/v_e$.

 In the next section I discuss the structure of each PN, 
and show that in most cases the PNs' structures can be used to strengthen 
the claims of Ciardullo {\it et al.} regarding the association, or
non-association, of the putative companions with the central stars of 
the PNs.

\section{NOTES ON THE INDIVIDUAL NEBULAE}

 In this section I analyze each PN separately.
 I will not discuss the position angle of the companion versus the
direction of the departure from axisymmetry.
 This is because there are no details on the orbit (e.g., velocity),
and the current position of the companion has little information
about the orbit inclination, eccentricity, etc.
 In analyzing the different PNs, I use images from Balick (1987),
Schwarz, Corradi, \& Melnick, (1992), Acker {\it et al.} (1992),
Manchado {\it et al.} (1996), and other sources
as indicated below. 

\noindent {\bf K1-14.}  
 There is a clear deviation from axisymmetry.
 The central star's distance from the center of the nebula $\Delta r$
is $\sim 15 \%$ of the nebular radius, i.e., 
$\Delta r / R_{PN} \simeq 0.15$.
 For an expansion velocity of 15 $\km \s^{-1}$ the age of the nebula
is $\sim 2 \times 10^4 \yrs$, just a little shorter than the orbital period.
 The velocity of the progenitor around the center of mass is relatively high,
$0.5 \km \s^{-1}$, and it can explain the departure from axisymmetry.
  Therefore, for this nebula the departure from axisymmetry may be attributed
to the presence of the companion.
  However, this is not the only possibility.
 The central star is closer to the north-east part of the nebula, where a
brighter arc is seen.
 This structure hints at an interaction with the ISM.
 Hence I put a question mark for this nebula (10th column in Table 1). 
 Future observations will have to examine in more detail the properties of
the companion and its association with the central star. 
 
\noindent {\bf A 63.} No clear deviation from axisymmetry is observed.
 This suggests that either the ``companion'' is not associated with the
progenitor of the nebula (Ciardullo {\it et al.} marked it as a
possible association), or that the real orbital separation is larger
than the projected orbital separation, hence the orbital period is
$\gtrsim 30$ times the age of the nebula.
 In any case, the absence of a clear departure from axisymmetry is compatible
with the properties of the orbit even if the observed companion is
a member of a binary system.

\noindent {\bf NGC 7008.} A clear departure is observed, both in the
outer and inner regions. This is expected, since the
mass-losing star's orbital velocity is large, $1.9 \km \s^{-1}$, and the
orbital period, although short, is within an order of magnitude of the
formation time of this small nebula.

\noindent {\bf NGC 650-51.} A clear asymmetry is seen along the waists
of this bipolar PN.
 It cannot be attributed to an interaction with the ISM.
 The orbital velocity of $\sim 0.26 \km \s^{-1}$ is too low
to explain the large departure in the equatorial plane. 
 The only way in which this ``companion'' could have caused the deviation
is that it has a very eccentric orbit and currently it is close to its
apastron position.
 In this case the orbital period is shorter, and when the companion was
closer the orbital velocity of the mass-losing star
was much higher.
 Although plausible, this is an unlikely explanation.
 More likely, a closer companion, but still at a wide orbital separation,
exists, or a close companion which directly affects the mass-loss process
orbits the central star with a highly eccentric orbit
(Soker {\it et al.} 1998).
 I marked this PN with a question mark.

 \noindent {\bf PuWe 1.} The images of this PN is from its discoverers,
Purgathofer \& Weinberger (1980), and Tweedy \& Kwitter (1996).
 This PN shows a clear departure from symmetry, but with a bright arc
on the outskirts of the nebula.
This clearly suggests an interaction with the ISM, as was argued by
Tweedy \& Kwitter (1996), rather than a wide binary companion influence.
 This is compatible with the claim of Ciardullo {\it et al.} that the
association of the observed ``companion'' with the central star is
doubtful. Hence I marked it with a +.
 
\noindent {\bf NGC 2392.} A small but clear departure is observed along the
south-north direction, which is the long axis of the nebula, in the outer
as well as the inner parts.
Along the minor axis, the eastern side is longer by several percents
than the western side.
 Such a deviation could not be caused by a companion at such a large orbital
separation.
 However, the distance found by Ciardullo {\it et al.}, $6.41 \kpc$, is an
upper limit.
 Most methods find a distance of $0.5-1.5 \kpc$ (Acker {\it et al.} 1992).
 Taking a distance to NGC 2392 to be $\sim 1.2 \kpc$, the orbital period is
$\sim 130,000 \yr$, the orbital velocity is $v_1 \sim 0.5 \km \s^{-2}$, and
the companion will be less massive.
 These parameters for the companion and orbital motion can explain the
observed departure from axisymmetry.
 Therefore, the morphology of this PN suggests that the companion
is associated with the central star.
 Hence I marked it with a + sign, despite the classification of
``possible association'' by Ciardullo {\it et al.}.
 Another possibility is discussed together with NGC 1535 below.

\noindent {\bf NGC 1535.} A small but clear asymmetry is observed along
position angle $\sim 220 ^\circ$.
 Such a small departure is compatible with the parameters of the
companion.
 Another possibility, for both this PN and NGC 2392, is that these are
elliptical PNs with their long axis almost along the line of sight.
 The two sides are not identical due to an unequal collimated flows
on the two sides of the long axis, which are not related to the presence
of the companion.
 A study of the expansion velocities of these PNs along different
directions is needed.

\noindent {\bf A 30.} This PN has a very small departure from sphericity
on its outskirts, which can safely be attributed to the ISM.
 No large-scale deviation is observed in the inner region,
only blobs and filaments,  hence I marked it to have no real deviation
from axisymmetry.
 This is compatible with the very long orbital motion
and low velocity.

\noindent {\bf A 7.} This a large nebula, which does not
show a bright arc, is located far from the galactic plane and is
surrounded by a tenuous ISM, hence it does not seem to interact with
the ISM (Xilouris {\it et al.} 1996; Tweedy \& Kwitter 1996).
 The distance to this PN is $\sim 0.2 \kpc$, much smaller than the upper
limit given by Ciardullo {\it et al.}, for which the orbital separation
is $\sim 200 \AU$ and the orbital period $\sim 2,500 \yr$.
 Such a binary should cause a noticeable departure.
 Although the nebula has an almost circular outer boundary, its
inner region is elliptical, with unidentical bright regions
on both sides.
 However, this is an old and large PN, and it is not clear whether
this small deviation from axisymmetry is related to the presence of
a companion, to instabilities which had time to grow, or to
a weak interaction with the ISM.
 Therefore, I marked it with a question mark.

\noindent {\bf A 24.} This PN has a point symmetric structure.
 The western side is a little more extended. I marked it
as having a questionable large-scale deviation from point symmetric.
 Although the orbital period is long, this is a large PN, hence
old one. 
 Tweedy \& Kwitter (1996) argue that it does not interact with the ISM.
 For all these, it is hard to tell whether its structure is compatible
with the companion properties.

\noindent {\bf A 31.} The image is taken from Tweedy \& Kwitter (1994),
who find that this PN interacts with the ISM.
 The morphology of this PN clearly comes from an interaction with the ISM.
Any influence by the companion will be much smaller than the effect
of the ISM.
 Therefore, I marked this PN with a question mark.

\noindent {\bf A33.} A large deviation from axisymmetry, which cannot
be attributed to the ISM since there is a deviation in the inner regions,
and no bright arc is observed on the side closer to the central star.
 The distance of the star from the center of the nebula is
$\Delta r/R_{PN} \simeq 0.07$.
 The departure of this PN is compatible with the properties of the companion,
as already noted in a previous paper (Soker 1997).

\noindent {\bf NGC 6210.} Ciardullo {\it et al.} find a distance smaller
than $ 32 \kpc$. Most other methods give distances of $\sim 2 \kpc$,
i.e., a factor of 16 smaller.
 This means an orbital period $\lesssim 64$ times shorter, and an
orbital velocity of $\lesssim 0.8 \km \s ^{-1}$.
 Such a binary companion should cause a noticeable deviation from axisymmetry.
 This PN shows a small deviation from axisymmetry along it major axis.
 Since the deviation is not so clear, and Ciardullo {\it et al.} argue for
a possible association between the companion and central star, I marked
it with a question mark.

\noindent {\bf NGC 3132.} This PN shows a clear asymmetry along
position angle $\sim 45 ^\circ$.
 In that direction the nebula extends to a distance
smaller by $\sim 20 \%$ than the distance in the opposite direction.
 Such a large departure from axisymmetry is expected from a companion
having the listed properties.

\noindent {\bf K 1-22.} Although the image in Acker {\it et al.} (1992)
is very faint, a small deviation from axisymmetry is observed in the
inner region, more or less along the minor axis of the PN.
This is compatible with the properties of the companion given in
Table 1.

\noindent {\bf K 1-27.} In Soker (1997) I suggested that the morphology
indicates an interaction with the ISM.
The finding of a companion by
Ciardullo {\it et al.} suggests that it is the companion that causes
the deviation from axisymmetry, or at least plays an important role
in it.
 For this PN Ciardullo {\it et al.} argue that the companion is a WD.
 Hence I took its mass to be $0.6 M_\odot$.

\noindent {\bf Mz 2.} A clear asymmetry is observed: the two outer lobes
are slightly displaced to the east relative to the inner bright shell, and
a ``missing'' segment in the shell is observed at position angle
$\sim 220 ^\circ$.
This departure is compatible with the finding of Ciardullo {\it et al.}.

\noindent {\bf Sp 3.}  In a previous paper (Soker 1997, Table 4)
 I noticed that the morphology of this PN suggests that it interacts
with the ISM or contains a wide binary companion.
 The findings of Ciardullo {\it et al.} show that it is the companion that
causes the departure from axisymmetry.
 
\noindent {\bf IC 4637.} A departure from axisymmetry is observed, but
it may come from an interaction with the ISM.
 I therefore marked it with a question mark. 

\section{PNs FOR WHOM NO COMPANIONS WERE FOUND}
{{{
   The purpose of the present paper is to demonstrate how the morphology of
a PN can be used to strengthen (or not) the claim for the presence
of a wide companion.
 However, in many cases a deviation from axisymmetry will be observed,
but no companion will be found.
 These cases are:
\newline
(1) The deviation is caused by the motion of the PN through the ISM.
I discussed this case in previous sections.
 Below a question arises in regard to A66 which has a clear asymmetry
(Hua, Dopita \& Mertinis 1998). However, this is a large low density
nebula, and therefore the asymmetry is most probably due to the ISM.
\newline
(2) The projected separation of the companion and the PN central star
is too small to be resolved, e.g.,
in the snapshot survey of Ciardullo {\it et al.}, the resolution
with the HST is $0.1$ arcsec. 
This will happen if the orbital separation is small, or if the orbital
separation is large but the binary plane is almost edge-on, and the
companion line of sight happens to be closed to that of the central star.
 Note that to observe a deviation from axisymmetry the orbit must not
be face on, so in many cases the orbit will be close to being edge-on.
\newline
(3) The companion is a low mass main sequence star, and it is too faint
to be detected.  In Table 1,  18 out of 19 companions, or possible
companions, have a mass of $\gtrsim 0.5 M_\odot$.
  Therefore, we can safely deduce that the study of Ciardullo {\it et al.}
is limited mainly to main sequence companions of masses
$\gtrsim 0.5 M_\odot$
(we return to this point below).
\newline
(4) The deviation from axisymmetry is due to a close companion at
an eccentric orbit (Soker {\it et al.} 1998). Such systems are likely
to form Bipolar PNs.
\newline
(5) A companion that enters the envelope of the progenitor of the PN
may cause a deviation from axisymmetry as well.
 This has been suggested as a possible explanation for the
displacement of the two outer rings of SN1987A (Soker 1999).
 Regarding mechanisms (4) and (5), a clear deviation from axisymmetry
is seen in the bipolar PN NGC 2346, for which there is a very close
companion with an orbital period of
16 days (Bond \& Livio 1990).
\newline
(6) In the survey of Ciardullo {\it et al.} PNs located in
a high field star density were also classified as non-detection,
since a wide binary companion could not be statistically identified
in these cases.
 It is possible therefore, that the companions of some of the
non-detection PNs which have nebular departure from axisymmetry
are seen in the survey, but can not be statistically distinguished from the
field stars. 

 With these possibilities in mind, I examine the PNs from the snapshot
survey of Ciardullo {\it et al.} for which no companions have been
detected.
 Here again, I consider a PN to have a clear departure from
axisymmetry when the scale of departure is larger than the
typical size of blobs and filaments in the PN.
The list of 67 PNs and their properties was kindly supplied to me by
R. Ciardullo.
 In this list there are 4 PNs which are classified as Bipolar,
and are likely to be formed from a close binary interaction
(Soker 1997, Table 3).
Considering point (4) above, I do not consider these PNs.
 Out of these 67 PNs, 13 PNs were claimed by Soker (1997, Table 4)
to be descendants of common envelope evolution.
Following point (5) above, I do not consider these PNs.

 Out of the rest 50 PNs, The deviations from axisymmetry of 15 PNs were
noted by Soker (1997) as follows. 
For 9 PNs the deviation from axisymmetry seems to result from 
interaction with the ISM;
two out of these 9 show no deviation in the inner regions, and therefore
have no influence of a wide companion (A16; NGC 6894), while 
for the other 7 it is hard to tell (due to, e.g., an extended filamentary
structure).
 For NGC 5979 Soker (1997) left it open whether the departure from
axisymmetry is due to an interaction with the ISM or to a wide companion.
IC 4593 show both signs of interaction with the ISM and a wide companion.
 For 4 PNs Soker (1997) attributes the departure to a wide companion
(IC 5148-50;NGC 2022; NGC 3242;  NGC 7662).
In these PNs the departure from axisymmetry is most prominent in
the inner regions, and therefore can be caused by a companion with
a relatively short orbital period.
 Soker, Zucker, \& Balcik  (1992) attributed the departure from axisymmetry
of NGC 3242 to a star with an orbital period of $\sim 4,000 \yrs$,
but a shorter period is also possible.
 A shorter orbital period is possible for NGC 2022, for which the
deviation is of the inner ``jets''.
 A careful examination of images in the literature reveals 
3 other PNs with a clear departure from axisymmetry (NGC 6153,
Gorny {\it et al.} 1999; NGC 5882, NGC 6804 Schwarz {\it et al.} 1992),
and 2 with departure which I cannot tell if compatible with
a wide companion (NGC 6891, Chu, Jacoby \& Arendt 1987;
NGC 2792, Schwarz {\it et al.} 1992).

 For 23 PNs out of the 50 I consider, I could not tell whether they have
a departure from axisymmetry which is compatible with a wide companion
(e.g., they have interaction with the ISM; they have filamentary structure
with small number of large filaments; no high resolution image exists).
 Out of the rest 27 PNs, 8 show clear departure, and for 3 it is not clear
(these 11 PNs were discussed in the previous paragraph).
 I conclude that $\sim 35 \pm 5 \%$ of PNs for whom no companions were
found by Ciardullo {\it et al.}, show departure from axisymmetry compatible
with the presence of a wide stellar companion.
  This should be compared to the 10 PNs with probable detection,
from which 8 have departure compatible with the claimed companion
(both 'Y' and '$+$' in Table 1).
 Considering the detection limits, and in particular the companion
mass of $\sim 0.5 M_\odot$, I find the differences in detection,
$80 \%$ compared with $35 \%$, significant, despite the small number
of PNs.

 I return now to the question of the limit on the companions masses.
 I used the I-band limiting magnitude (supplied by Ciardullo)
of the 8 PNs I claim to have wide companions and the 3 PNs
with question marks,
to estimate the upper limit on the masses of possible main sequence
companions.
 I took distances from Cahn, Kaler \& Stanghellini (1992), and the
mass-luminosity relation for main sequence stars.
 Because of the strong dependence of luminosity on mass in that relation,
the calculation is quite robust to the uncertainties in the different
quantities entering the calculation.
 The uncertainties in the distances may be up to a factor of $\sim 3$,
which introduce uncertainties of up to $\sim 70 \%$ in the limiting masses.
 Because of these uncertainties I do not find it appropriate
to list each object separately (unlike the case of the PNs presented
in Table 1, for which Ciardullo {\it et al.} used the companions to
determine the distances).
 Overall, I find the limiting masses to be around $\sim 0.5 M_\odot$
for most objects.
 This is compatible with the masses of detected companions
(Table 1).
 
 A detailed statistical study of binary systems that cause departure
from axisymmetry using a population synthesis code, is planned for
a forthcoming paper.
 The code that will be used is the one used by Rappaport \& Soker (1999),
from which I take the different values used below.
 I crudely estimate the number of expected wide binaries as follows.
For an orbital separation smaller than $\sim 1 \AU$ (an orbital period of
one year), a strong interaction of the binary system will form a bipolar
PNs, or the system will enter a common envelope.
I exclude such systems here.
 For the rest, the distribution of binary orbital period is
uniform in $\log (P)$, up to $P \simeq 10^6 \yrs$. 
 The orbital velocity of the mass-losing star, of mass $M_1$,
around the center of mass with a companion of mass $M_2$, is
(assuming a circular orbit)
\begin{eqnarray}
v_1= 0.4
\left( {{P} \over {10^4 \yr}} \right)^{-1/3}
\left( {{M_1+M_2} \over {1 M_\odot}} \right)^{-2/3}
\left( {{M_2} \over {0.3 M_\odot}} \right)
\km \s^{-1}.
\end{eqnarray}
 As mentioned before, an orbital velocity of $\sim 0.3 \km \s^{-1}$
may cause a noticeable departure from axisymmetry.
 The mass distribution in binary systems is somewhat peaked toward
$M_2/M_1 =1$, hence most companions of PN progenitors
will have masses of $M_2 \gtrsim 0.3 M_\odot$.
  This means that binaries with orbital periods of up to
few$\times 10^5 \yrs$ can be counted, especially if eccentric orbits
are considered as well. 
 From below, we require that the orbital period will not bee too short,
this gives an orbital period longer than $\sim 500 \yrs$.
 Overall, most companions in a logarithmic interval of $\sim 2$,
which is $\sim 1/3$ of all binary systems may cause deviation from
axisymmetry.
 The question is what is the fraction of binary systems among
 all progenitors of PNs. This fraction is taken by different authors to
 be in the range of 0.6-1 (Rappaport \& Soker 1999).
  I conclude that wide binaries will cause a clear departure from axisymmetry
in $20-35 \%$ of all PNs.
 The fraction of PNs below the detection limit mentioned above,
 I crudely estimate to be half this number $10-17 \%$.
 Adding to these some interaction with the ISM (e.g., a dense cloud),
and large scale instabilities in the mass loss process,
which I interpreted as wide binaries in analyzing the PNs in this section,
I find the fraction of PNs with departure from pure axisymmetry
among the non-detection PNs ($\sim 35 \%$)
to be quite reasonable.
}}}

\section{SUMMARY}

 Stellar companions to progenitors of PNs will influence the morphology
of the descendant PNs. Close companions will affect directly the
mass loss process from the AGB progenitor. Such close binaries
are not likely to be resolved.
 Very wide binaries (as well as the wide binaries discussed here) may form a
small bubble inside the nebula (Soker 1996).
  In the intermediate range, binary systems which have orbital periods
in the range of several$\times 100$ to a few$\times 10^5$ years, may
cause the PN to have a large-scale departure from axisymmetrical structure.
  The goal of the present paper was to show that the structure of a
PN can be used to strengthen an argument for an association, or a
non-association, of a putative wide companion with the progenitor of
the PN.
In a recent work Ciardullo {\it et al.} (1999) list 19 PNs for
which they argue a probable, possible, or doubtful association of
stars with the progenitors of the PNs.
 Ciardullo {\it et al.} did not use the structure of the PNs as a tool to
further support their claims.
 In the present paper I used their list, and demonstrated that the
structures of the PNs are compatible in most cases with their claims,
hence this paper strengthens them.
 The greatest confusion may come from a possible influence of the ISM
on the PN morphology, especially for large PNs.
 I tried to present arguments for and against ISM interaction where possible.

 In the opposite direction, PNs whose central stars have companions
can be used to further study the influence of wide companions
on the PNs morphology.
 For this, a determination of the binary component velocity is needed.
 Very helpful but more difficult to determine will be the eccentricity
and inclination of the orbital plane.

{{{  From a list of 27 PNs observed with HST for which no companions
were found by Ciardullo {\it et al.}, I claimed that 8 PNs are likely 
to have companions which have caused their structures to depart 
from axisymmetry.
 In addition to 5 PNs from this list which were predicted to have wide
binary companions by Soker (1997), I added here to the list three PNs:
NGC 5882, NGC 6153, and NGC 6804.
 I estimated that the observations of Ciardullo {\it et al.}
will not detect main sequence companions of masses below
$\sim 0.5 M_\odot$ in these 8 PNs. 
  A search for a wide binary companion in each of these 8 PNs is encouraged.
 For three PNs out of the 27 I could not tell if a companion is compatible
or not with their structure. 
 That only 8 of the 27 PNs have morphology compatible with a wide binary
companion, compared with 8 out of 10 PNs that were claimed to have probable
companion by Ciardullo {\it et al.}, strengthen the arguments
presented in the present paper. }}}

 {\bf ACKNOWLEDGMENTS:} 
{{{I would like to thank the referee, Robin Ciardullo,
for his suggestion to study the sample of PNs for whom no companions
were found, and for supplying the data for these PNs,
and Matthew Bobrowsky for a careful reading of
the manuscript.  }}}
This research was supported in part by a grant from the University of
Haifa and a grant from the Israel Science Foundation.

\end{document}